\documentclass[pra,showpacs,superscriptadress]{revtex4}

\usepackage{graphicx}
\usepackage{amsmath}
\usepackage{epsfig}
\usepackage{bm}

\begin{document}

\title{Hong-Ou-Mandel interferometer with cavities: theory}
\author{C. Olindo }
\author{M. A. Sagioro}
\author{C. H. Monken}
\author{S. P\'{a}dua}
\affiliation{Departamento de F\'{\i}sica, Universidade Federal de
Minas Gerais, Caixa Postal 702, 30123-970 Belo~Horizonte~MG,
Brazil.}
\author{A. Delgado}
\affiliation{Departmento de F\'{\i}sica\\ Centro de Optica
Cu\'antica e Informaci\'on\\ Universidad de Concepci\'on, Chile}
\date{\today}

\begin{abstract}
We study the number of coincidences in a Hong-Ou-Mandel
interferometer exit whose arms have been supplemented with the
addition of one or two optical cavities. The fourth-order
correlation function at the beam-splitter exit is calculated. In
the regime where the cavity length are larger than the one-photon
coherence length, photon coalescence and anti-coalescence
interference is observed. Feynman's path diagrams for the
indistinguishable processes that lead to quantum interference are
presented. As application for the Hong-Ou-Mandel interferometer
with two cavities, it is discussed the construction of an optical
XOR gate.
\end{abstract}

\maketitle

\section{Introduction}
\label{intro}

Spontaneous parametric down-conversion (SPDC) is an important
resource for quantum optics studies. In the process of SPDC, a
pump (\it{p}\rm) laser beam incident upon a nonlinear crystal
creates a pair of entangled photons, usually called signal
(\it{s}\rm) and idler (\it{i}\rm) \cite{mandel}. SPDC was first
proposed theoretically by Klyshko in 1969 \cite{klyshko} and
demonstrated experimentally by Burnham and Weinberg in 1970
\cite{burnham}. Idler and signal quantum interference in a
beam-splitter was first demonstrated by Hong, Ou and Mandel (HOM)
in 1987 \cite{HOM}. In their experiment, signal and idler photons
with the same frequency and polarization are combined in a 50/50
beam-splitter (BS) and the output photons are detected at the exit
of the beam-splitter by coincidence detection. When the idler and
the signal paths are made equal from the crystal to the BS, no
coincidence counts are detected at the BS output. This null result
is due to a destructive interference between the two
indistinguishable paths that the pair can follow to produce
coincidences: both photons are reflected at the BS or both photons
are transmitted. Since there is a $\pi/2$ phase shift for the
photon reflection probability amplitude at the BS, the probability
amplitudes for these two indistinguishable events cancel out and
zero coincidence or a minimum, is measured \cite{steinberg}.
Therefore, at this point the photon-pair goes to either exit
(photon coalescence \cite{Wang}).

Optical cavities have been used in SPDC quantum optics experiments
recently aiming the demonstration of the control of photon
coalescence and anti-coalescence via a cavity and the production
of time entangled photon-pairs. Sagioro, Olindo, Monken and
P\'{a}dua \cite{Sagioro}, have demonstrated experimentally the
interference of photons with the same polarization generated by
spontaneous parametric down-conversion in a Hong-Ou-Mandel
interferometer, after one of them passed through a symmetric
cavity, i.e. a cavity whose mirrors have equal reflectivity. In
the regime where the cavity length is larger than the one-photon
coherence length, photon coalescence and anti-coalescence
interference is observed. It is shown that by changing the cavity
length, coincidence peaks can be transformed in dips and
vice-versa, even though the coherence length of the photon
wavepacket is smaller than the cavity length.  With a different
goal, Lu, Campbell and Ou \cite{ou}, have generated mode-locked
two-photon states by using a Fabry-P\`{e}rot cavity for filtering
the wide band light produced by SPDC. The comb-like time
correlation of the photon pairs is observed with a HOM
interferometer at the exit of the Fabry-Perot cavity. Zavatta,
Viciani and Bellini \cite{zavatta}, have generated comb-like
two-photon entangled states by inserting in one of the arms of a
HOM interferometer a Fabry-Perot cavity. In this manuscript, we
analyze theoretically a modified version of the Hong-Ou-Mandel
interferometer. The optical paths along the arms of the
interferometer contain two Fabry-Perot cavities, each one in an
arm of the interferometer. These cavities allow the partial
transmission and reflection of photons on each arm leading to a
deviation of the number $N_c$ of coincidences from the standard
Hong-Ou-Mandel interferometer.

\section{Experimental setup}
%MODIFICAR O Desenho DA FIGURA.
\begin{figure}[h]
\centerline{\includegraphics[angle=-90,width=8cm]{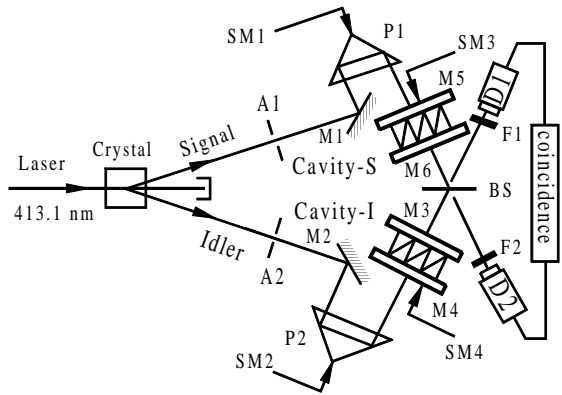}}
\caption{Experimental scheme of a Hong-Ou-Mandel interferometer in
which two Fabry-P\'{e}rot cavities are inserted in the arms of the
interferometer. The symbols $BS$, $M$, $P$, $SM$, $A$, and $D$
label beam splitters, mirrors, prisms, step motors, apertures, and
detectors, respectively.} \label{hom2}
\end{figure}

The experimental apparatus, represented in Fig. (\ref{hom2}),
consists of an HOM interferometer \cite{HOM}, and two symmetric
cavities formed by two planar dielectric mirrors M5 and M6
(M3 and M4). A $5$ mm long ${\rm
LiIO}_{3}$ crystal, oriented for type-I phase matching, is pumped
by a CW laser oscillating at $413.1$ nm. A violet photon from the
pump beam is down-converted into two conjugate infrared photons at
$826.2$ nm (signal and idler), so that energy and momentum are
conserved during this process. Both signal and idler photons are
horizontally polarized and propagate at some angle $\theta$ from
the pump beam direction. The apertures A1, A2 select out the
desired degenerate photon pairs. Both beams are directed by the
prisms P1, P2, and mirrors M1, M2 into the two input ports of the
$50/50$ beam splitter BS. Each beam pass through symmetric
Fabry-Perot cavities. Both detectors
D1 and D2 are avalanche photodiodes operating in photon counting
mode and F1, F2 are equal interference filters centered at $826.2$
nm. The prism P1 (P2) is fixed over a translation stage
displaceable by a step motor SM1 (SM2) used to change the signal
(idler) path length. The step motor SM3 (SM4) is used for moving
the mirror M5 (M4) for changing the cavity length
$L$.

\section{Analytical results}
We consider that the photon-pairs are generated not collinearly
(Fig.\,\ref{hom2}) by SPDC process with the crystal cut for type I
phase-matching. In this case, signal and idler photons are
generated with the same linear polarization. Suppose, we select
the idler and signal photons with the same frequency and equal to
half of the frequency of the laser pump beam (degenerate case):
$\omega_{i}=\omega_{s}=\omega_{p}/2$. We also assume that the
photon beams are selected by small diameter apertures (see
Fig.\,\ref{hom2}) such that the propagation directions are well
selected. In this case, the wavevector dispersion can be neglected
and the momentum conservation in the SPDC process considered
perfect. We also consider that the pump laser beam is a coherent
monochromatic beam, with no dispersion in frequency. The photons
are then described by the state \cite{mandel,HOM}
\begin{equation}
|\Psi\rangle=\int_0^{\omega_p} d\omega\phi(\omega)
|1,\omega\rangle_i|1,\omega_p-\omega\rangle_s. \label{SPDC1}
\end{equation}
where the function $\phi(\omega)$ is centered in $\omega_p/2$. In
practice $\phi(\omega)$  is defined by interference filters in
front the detectors (see Fig.\,\ref{hom2}). The state
$|1,\omega\rangle_i$ ($|1,\omega_p-\omega\rangle_s$) describes one
photon of frequency $\omega$ ($\omega_p-\omega$) in the idler
(signal) beam. The contribution from the vacuum cancels out in
a photo-coincidence experiment as we have described above, so its
contribution have been neglected in Eq. (\ref{SPDC1}). The state
$|\Psi\rangle$ can also be cast in the form
\begin{equation}
|\Psi\rangle=\int_0^{\omega_p}d\omega\phi(\omega)
a_{i}^{\dagger}(\omega)a_{s}^{\dagger}(\omega_p-\omega)|\rm{vacuum}\rangle,
\label{SPDC2}
\end{equation}
where the operators $a_{i}^{\dagger}(\omega)$ and
$a_{s}^{\dagger}(\omega_p-\omega)$ create photons of frequency
$\omega$ and $\omega_p-\omega$ in the idler and signal modes,
respectively.

After being generated, the photons of each beam interact with a
Fabry-Perot cavity. The photons can be transmitted or reflected by
the cavities. The action of these cavities is similar to a beam
splitter \cite{monken}, that is
\begin{eqnarray}
a_{i}^{\dagger}(\omega)&=&\mu_{i}(\omega)a_{i,t}^{\dagger}(\omega)+
\nu_{i}(\omega)a_{i,r}^{\dagger}(\omega),
\label{Transformation1}\\
a_{s}^{\dagger}(\omega)&=&\mu_{s}(\omega)a_{s,t}^{\dagger}(\omega)+
\nu_{s}(\omega)a_{s,r}^{\dagger}(\omega),
\label{Transformation2}
\end{eqnarray}
where the operators $a_{i,t}^{\dagger}(\omega)$ and
$a_{i,r}^{\dagger}(\omega)$ ($a_{s,t}^{\dagger}(\omega)$ and
$a_{s,r}^{\dagger}(\omega)$) describe reflected and transmitted
photons of the incoming idler (signal) beam with frequency
$\omega$. The coefficients $\mu_{i}(\omega)$, $\nu_{i}(\omega)$,
$\mu_{s}(\omega)$, and $\nu_{s}(\omega)$ are the cavities
transmission and reflection coefficients for idler and signal beam,
respectively. Thereby, the state $|\tilde\Psi\rangle$ of the beams
after the interaction with the cavities is given by the expression
\begin{eqnarray}
|\tilde\Psi\rangle &= &\int_0^{\omega_p}d\omega\phi(\omega)
 \big(\mu_{i}(\omega)\mu_{s}(\omega_p-\omega)|1,\omega\rangle_{i,t}
|1,\omega_p-\omega\rangle_{s,t}
+\mu_{i}(\omega)\nu_{s}(\omega_p-\omega)
|1,\omega\rangle_{i,t}|1,\omega_p-\omega\rangle_{s,r}
\nonumber\\
 & &\quad\quad \quad \quad \quad \, + \nu_{i}(\omega)\mu_{s}(\omega_p-\omega)
|1,\omega\rangle_{i,r}|1,\omega_p-\omega\rangle_{s,t}
+\nu_{i}(\omega)\nu_{s}(\omega_p-\omega) |1,\omega\rangle_{i,r}
|1,\omega_p-\omega\rangle_{s,r}\big). \label{Aftercavities}
\end{eqnarray}
The transmitted beams through the cavities are directed to a
$50/50$ beam splitter (BS). The two output beams are then
detected in coincidence at the BS exit ports. The state of the
beams after the cavities, Eq. (\ref{Aftercavities}), is a
superposition of four mutually orthogonal states. These states
describe the reflection or transmission of photons in each beam by
the cavities. Consequently, only the transmitted photons of both
beams can be detected at the exit of BS.

The field operators at the detectors of the interferometer are
\cite{HOM}
\begin{eqnarray}
E_{1}^{+}(t)&=&\frac{1}{\sqrt{2}}[E_{i,t}^{+}(t-\tau_1-\delta_i)
+iE_{s,t}^{+}(t-\tau_1-\delta_s)],
\nonumber\\
E_{2}^{+}(t)&=&\frac{1}{\sqrt{2}}[E_{s,t}^{+}(t-\tau_1-\delta_s)
+iE_{i,t}^{+}(t-\tau_1-\delta_i)],
\end{eqnarray}
where $\tau_1$ is the propagation time from mirrors to detectors
and $\delta_i$ and $\delta_s$ are the time delays in the idler and
signal beams due to a modification of their optical paths (P1 and
P2 can be displaced, see Fig.\,\ref{hom2}).

The field operators at the BS input are given by
\begin{eqnarray}
E_{i,t}^{+}(t)&=&\int d\omega e^{-i\omega t}a_{i,t}(\omega),
\nonumber\\
E_{s,t}^{+}(t)&=&\int d\omega e^{-i\omega t}a_{s,t}(\omega).
\end{eqnarray}

The probability $P(\tau)$ of detecting one photon on a detector in
the time $t$ and a second photon in the other detector, in the
time $t+\tau$ is
\begin{equation}
P(\tau)=Tr(\rho
E_{1}^{-}(t)E_{2}^{-}(t+\tau)E_{2}^{+}(t+\tau)E_{1}^{+}(t)),
\end{equation}
where the density operator $\rho$ describes the state of the electromagnetic field
before the beam splitter \cite{mandel}. The number $N_c$ of coincidences can be
obtained by integrating this probability in the time interval $\tau$ of the coincidence window.

% Let us calculate the number $N_c$ of coincidences at the output of BS to be
% detected when both beams are described by the state shown in Eq.
% (\ref{Aftercavities}). The state $|\tilde\Psi\rangle$ can be cast
% as
% $|\tilde\Psi\rangle=(1/2)(|\tilde\Psi\rangle_1+|\tilde\Psi\rangle_2)$,
% where the states $|\tilde\Psi\rangle_1$ and $|\tilde\Psi\rangle_2$
% are given by
% \begin{eqnarray}
% |\tilde\Psi_1\rangle&=&\int_0^{\omega_p}d\omega\phi(\omega)
% \Big(\mu_{i,\omega}\mu_{s,\omega_p-\omega}|1,\omega\rangle_{i,t}|1,\omega_p-\omega\rangle_{s,t}+
% \nu_{i,\omega}\mu_{s,\omega_p-\omega}|1,\omega\rangle_{i,r}|1,\omega_p-\omega\rangle_{s,t}\Big)
% \nonumber\\
% |\tilde\Psi_2\rangle&=&\int_0^{\omega_p}d\omega\phi(\omega)\Big
% (\mu_{i,\omega}\nu_{s,\omega_p-\omega}|1,\omega\rangle_{i,t}|1,\omega_p-\omega\rangle_{s,r}+
% \nu_{i,\omega}\nu_{s,\omega_p-\omega}|1,\omega\rangle_{i,r}|1,\omega_p-\omega\rangle_{s,r}\Big).
% \end{eqnarray}
% The second state $|\tilde\Psi_2\rangle$ does not contributes in the calculations of the
% field amplitudes since it describes photons of the signal beam
% which are reflected by the cavity. Clearly, these photons cannot
% lead to coincidences in the detectors at the output of the HOM interferometer. A similar reason show us
% that only the first term in $|\tilde\Psi_1\rangle$ contributes to
% the number of coincidences.

Due to the purity of the two-photon state, the probability $P(\tau)$ can be calculated as
\cite{mandel}
\begin{equation}
P(\tau)=\langle\Psi_1|E_{1}^{-}(t)E_{2}^{-}(t+\tau)E_{2}^{+}(t+\tau)
E_{1}^{+}(t)|\Psi_1\rangle,
\label{one}
\end{equation}
where the state
\begin{equation}
|\Psi_1\rangle=\int_0^{\omega_p}d\omega\phi(\omega)
\mu_{i,\omega}\mu_{s,\omega_p-\omega}|1,\omega\rangle_{i,t}|1,\omega_p-\omega\rangle_{s,t}
\end{equation}
describes photons transmitted through both cavities.

Let us now consider the state $E_{2}^{+}(t+\tau)E_{1}^{+}(t)|\Psi_1\rangle$.
Thereafter the probability $P(\tau)$ can be calculated, according to Eq. (\ref{one}), as
the square of the norm
of this state. The action of the operator $E_{2}^{+}(t+\tau)E_{1}^{+}(t)$ onto the state
$|\Psi_1\rangle$ is
\begin{eqnarray}
E_{2}^{+}(t+\tau)
E_{1}^{+}(t)|\Psi_1\rangle&=&
\frac{1}{2}(E_{s,t}^{+}(t+\tau-\tau_1-\delta_s)E_{i,t}^{+}(t-\tau_1-\delta_i)
|\Psi_1\rangle
\nonumber\\
&+&\frac{i}{2}(E_{s,t}^{+}(t+\tau-\tau_1-\delta_s)E_{s,t}^{+}(t-\tau_1-\delta_s)
|\Psi_1\rangle
\nonumber\\
&+&\frac{i}{2}(E_{i,t}^{+}(t+\tau-\tau_1-\delta_i)E_{i,t}^{+}(t-\tau_1-\delta_i)
|\Psi_1\rangle
\nonumber\\
&-&\frac{1}{2}(E_{i,t}^{+}(t+\tau-\tau_1-\delta_i)E_{s,t}^{+}(t-\tau_1-\delta_s)
|\Psi_1\rangle. \label{E2E1psi1}
\end{eqnarray}
The first term at the r.h.s. of the previous equation can be cast in the form

\begin{equation}
\frac{1}{2}\int\int\int d\omega d\omega' d\omega''
e^{-i\omega'(t+\tau-\tau_1-\delta_s)}
e^{-i\omega''(t-\tau_1-\delta_i)} \phi(\omega)
\mu_{i}(\omega)\mu_{s}(\omega_p-\omega)a_{i,t}(\omega'')a_{i,t}^{\dagger}(\omega)
a_{s,t}(\omega')a_{s,t}^{\dagger}(\omega_p-\omega)
%\nonumber\\
 |\rm{vacuum}\rangle. \label{E2E1psiprino}
\end{equation}

 The operators $a_{i,t}(\omega'')$ and $a_{s,t}(\omega')$
entering in Eq. (\ref{E2E1psiprino}) destroys the vacuum state
unless the conditions $\omega''=\omega$ and
$\omega'=\omega_p-\omega$ hold. In this case we obtain
\begin{equation}
\frac{1}{2}(E_{s,t}^{+}(t+\tau-\tau_1-\delta_s)E_{i,t}^{+}(t-\tau_1-\delta_i)
|\Psi_1\rangle= \frac{1}{2}\int d\omega
e^{-i(\omega_p-\omega)(t+\tau-\tau_1-\delta_s)}
 e^{-i\omega(t-\tau_1-\delta_i)}\phi(\omega) \mu_{i}(\omega)\mu_{s}(\omega_p-\omega)|\rm{vacuum}\rangle.
\label{E2E1si1}
\end{equation}
In a similar way it can be show that
\begin{equation}
\frac{1}{2}(E_{i,t}^{+}(t+\tau-\tau_1-\delta_i)E_{s,t}^{+}(t-\tau_1-\delta_s)
|\Psi_1\rangle=
\frac{1}{2}\int d\omega e^{-i\omega(t+\tau-\tau_1-\delta_i)}
 e^{-i(\omega_p-\omega)(t-\tau_1-\delta_s)}\phi(\omega) \mu_{i}(\omega)\mu_{s}(\omega_p-\omega)|\rm{vacuum}\rangle,
\label{E2E1si2}
\end{equation}
being the remaining terms entering in Eq. (\ref{E2E1psi1}) zero.
Thereby, we have
\begin{eqnarray}
E_{2}^{+}(t+\tau)E_{1}^{+}(t)|\Psi_1\rangle=
\frac{1}{2}e^{-i\omega_p(t-\tau_1-\delta_s)} \lbrace
&e^{-i\omega_p\tau}&\int_0^{\omega_p} d\omega
e^{i\omega\tau}e^{i\omega(\delta_i-\delta_s)}
\phi(\omega)\mu_{i}(\omega)\mu_{s}(\omega_p-\omega)
\nonumber\\
&-&\int_0^{\omega_p} d\omega
e^{-i\omega\tau}e^{i\omega(\delta_i-\delta_s)}
\phi(\omega)\mu_{i}(\omega)\mu_{s}(\omega_p-\omega)
\rbrace|\rm{vacuum}\rangle. \label{E2E1sito}
\end{eqnarray}
A further simplification of this expression can be obtained by a
variable change and a global phase omission. We obtain for Eq.
(\ref{E2E1psi1})
\begin{equation}
E_{2}^{+}(t+\tau)E_{1}^{+}(t)|\Psi_1\rangle=
\int_{-\frac{\omega_p}{2}}^{\frac{\omega_p}{2}} d\omega(e^{i\tau}-e^{-i\tau})
e^{i\omega(\delta_i-\delta_s)} \phi(\omega+\frac{\omega_p}{2})
\mu_{i}(\omega+\frac{\omega_p}{2})\mu_{s}(\frac{\omega_p}{2}-\omega)|\rm{vacuum}\rangle.
\end{equation}
The transmission coefficients $\mu_s$ and $\mu_i$ entering in this state are given by
\begin{equation}
\mu_{j}(\omega)=t_{j}^{2}\sum_{m=0}^{\infty}r_{j}^{2m}e^{2im\omega\tau_{cj}}
\label{Transmission_coefficient}
\end{equation}
where $j = i,s$ denotes the $idler$ or $signal$ beam. $t_{j}$ and $r_{j}$ are
the transmission and reflection coefficients of the cavity respectively and
$\tau_{cj} = L_{j}/c$ with $L_{j}$ being the $j$ cavity length and $c$ the light velocity.
This particular form of the $\mu_j$ coefficient arises from considering the cavities as two
 planar dielectric mirrors with the same reflection and transmission coefficients
 ($r_j$ and $t_j$) separated by a distance $L_j$. Each time a monochromatic plane wave of
  frequency $\omega$ enters the cavity there are multiple reflections and transmissions of
  the field on the mirrors. Each time the electromagnetic field of the wave cross a mirror
  of the cavity, the field is multiplied by $t_j$ and each time it is reflected, it is
   multiplied by $r_j$. We must also consider the field propagation between the mirrors
   (here is the dependence of $\mu_j$ on the product of $\omega$ and $L_j$). Finally,
    adding all the contributions of the electromagnetic fields that are transmitted by the
     cavity we get $E_t=\mu E_{0}$, where $E_{t}$ is the transmitted field and $E_0$ the
     incident one.

The probability $P(\tau)$ is then given by
\begin{eqnarray}
P(\tau)&=&\frac{K(T_{i} T_{s})^{2}}{4}\sum_{m,n,l,q} R_{s}^{m+l}
R_{i}^{n+q} e^{i\omega_p\tau_{cs}(m-l)}e^{i\omega_p\tau_{ci}(n-q)}
\int_{-\frac{\omega_p}{2}}^{\frac{\omega_p}{2}}
\int_{-\frac{\omega_p}{2}}^{\frac{\omega_p}{2}} d\omega d\omega'
e^{i\omega(2m\tau_{cs}-2n\tau_{ci}-\delta)}
e^{-i\omega'(2l\tau_{cs}-2q\tau_{ci}-\delta)}
\nonumber\\
&&\phi(\omega+\frac{\omega_p}{2})\phi^{*}(\omega'+\frac{\omega_p}{2})
\Big( e^{i\tau(\omega-\omega')}-e^{i\tau(\omega+\omega')}
-e^{-i\tau(\omega+\omega')}+e^{-i\tau(\omega-\omega')} \Big),
\end{eqnarray}
where $\delta = \delta_i-\delta_s$ is the difference of the optical paths of both beams,
$T_{j}=|t_{j}|^2$ and $R_{j}=|r_{j}|^2$ are the transmittance and reflectance of the $j$
 cavity respectively and $K$ is proportional to the detectors efficiency.

After integrating over $\tau$ and assuming symmetry of $\phi(\omega+\frac{\omega_p}{2})$
around $\frac{\omega_p}{2}$, the number $N_c$ of coincidences becomes
\begin{eqnarray}
N_c=\pi K (T_{i} T_{s})^{2}&&\sum_{m,n,l,q} R_{s}^{m+l}
R_{i}^{n+q}
\cos(\omega_p\tau_{cs}(m-l))\cos(\omega_p\tau_{ci}(n-q))
\int_{-\frac{\omega_p}{2}}^{\frac{\omega_p}{2}} d\omega
\nonumber\\
&& \Big[ \cos\big(2\omega(\tau_{cs}(m-l)-\tau_{ci}(n-q))\big) -
\cos\big(2\omega(\tau_{cs}(m+l)-\tau_{ci}(n+q))-\delta\big) \Big]
|\phi(\omega+\frac{\omega_p}{2})|^{2}.
\end{eqnarray}
A further simplification can be achieved considering a gaussian
distribution of width $\bigtriangleup\omega$ for the function
$\phi(\omega+\frac{\omega_p}{2})$, supposed sufficiently narrow in
the interval $(-\frac{\omega_p}{2},\frac{\omega_p}{2})$ such that
the limits of integration can be extended to $(-\infty,\infty)$.
Thereby, we finally obtain for the number $N_c$ of coincidences
the expression
\begin{eqnarray}
N_c=\pi K (T_{i} T_{s})^{2}&&\sum_{m,n,l,q} R_{s}^{m+l}
R_{i}^{n+q}
\cos(\omega_p\tau_{cs}(m-l))\cos(\omega_p\tau_{ci}(n-q))
\nonumber\\
&& \Big[
e^{\big(-\bigtriangleup\omega^{2}(\tau_{cs}(m-l)-\tau_{ci}(n-q))^{2}\big)}
-e^{\big(-\bigtriangleup\omega^{2}
(\tau_{cs}(m+l)-\tau_{ci}(n+q)-\delta)^{2}\big)} \Big].
\label{eqg}
\end{eqnarray}
%%%%%%%%%parei

\section{HOM with one cavity}

Here we study interferometric properties of HOM interferometer
with one cavity. To obtain the number of coincidences $N_c$ in
this particular case we make $R_{s}=0$, $T_{s}=1$ and
$\tau_{cs}=0$ in Eq. (\ref{eqg}). Thereby, we consider a single
cavity in the path of the idler beam. The number of coincidences
becomes
\begin{equation}
N_c= T^2\sum_{n,q}
R^{n+q}
\cos\big(\omega_p\tau_c(n-q)\big)
 \Big(
e^{-\bigtriangleup\omega^{2}\tau_c^2(n-q)^2}
-e^{-\bigtriangleup\omega^2
(\tau_c(n+q)-\delta)^2} \Big),
\label{equ}
\end{equation}
with $T=T_{i}$, $R=R_{i}$, $\tau_c=\tau_{ci}$ and the constant
$\pi K$ was removed because we are interested in the coincidence
rate, not in the number of coincidences itself. In particular,
making $T=1$, $R=0$ and $\tau_c=0$ we recover from Eq. (\ref{equ})
the well known expression for the number of coincidences of the
HOM interferometer without cavities, that is
\begin{equation}
N_c=1-e^{-\Delta\omega^2\delta^2}.
\end{equation}
Before proceeding with the analysis of Eq. (\ref{equ}) we make a change of variables.
This equation becomes
\begin{equation}
N_c= T^2\sum_{n,q} R^{n+q} \cos\big(\frac{2\pi
L}{\lambda_p}(n-q)\big) e^{-(\frac{2\pi \bigtriangleup\lambda
L}{\lambda^2}(n-q))^2} - T^2\sum_{n,q} R^{n+q} \cos\big(\frac{2\pi
L}{\lambda_p}(n-q)\big) e^{-(\frac{2\pi c
\bigtriangleup\lambda}{\lambda^2})^2 (\frac{L}{c}(n+q)-\delta)^2}
\label{equ1}
\end{equation}
where $\lambda$ is the central wavelength of the twin photons
wavepackets, $\lambda_p$ is the pump wavelength,
$\bigtriangleup\lambda$ characterizes the filter in front of the
detectors, $c$ is the speed of light and $L$ is the length of the
cavity.

\begin{figure}[h]
\centerline{\includegraphics[angle=0,width=8cm]{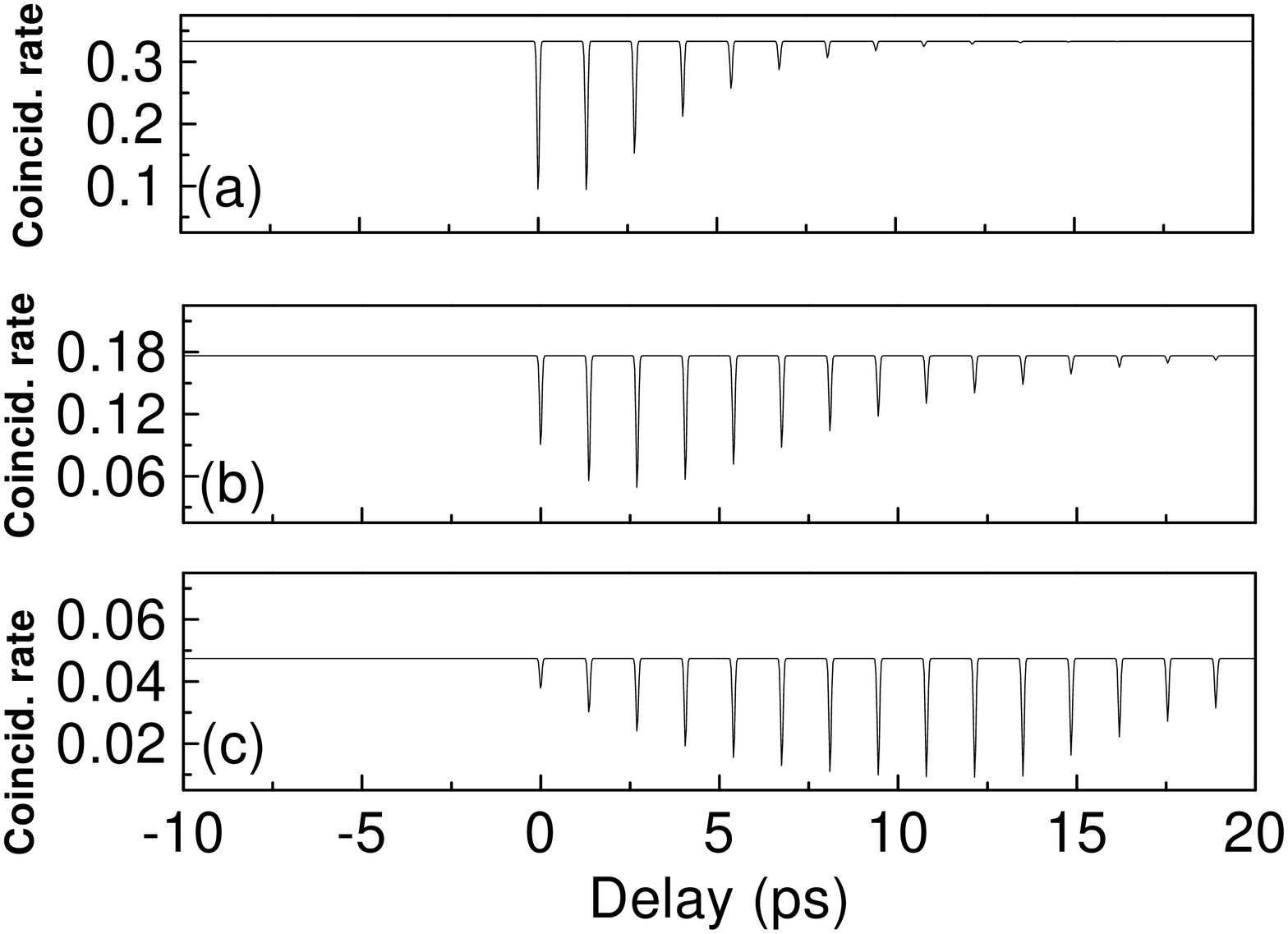}}
\caption{In all three graphics the wavelength of the degenerate
photons ($\lambda$) was $826.2$ nm and the interference filters in
front of the detectors had bandwidth $(\Delta\lambda)$
equal to $8$ nm. In
all three cases $L$ was equal to $0.404838$ mm (to achieve the resonant situation).
In (a) $R=0.5$, in (b) $R=0.7$, in (c)$ R=0.9$ and $R+T=1$ (no losses).}
\label{ucvr}
\end{figure}

\begin{figure}[h]
\centerline{\includegraphics[angle=0,width=8cm]{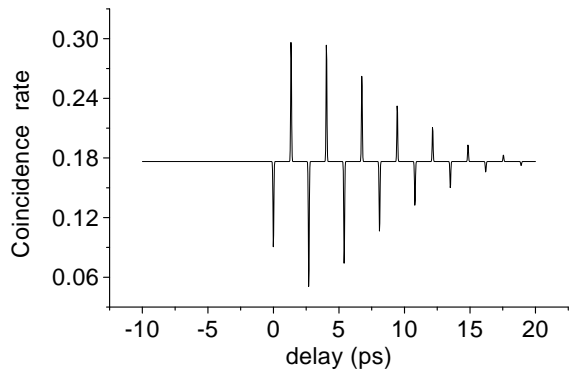}}
\caption{The setup parameters are: $\lambda=826.2$ nm, $\Delta\lambda=8$ nm, $R=0.7$ and
$L=0.4050447$ mm.} \label{ucar}
\end{figure}

\begin{figure}[h]
\centerline{\includegraphics[angle=0,width=10cm]{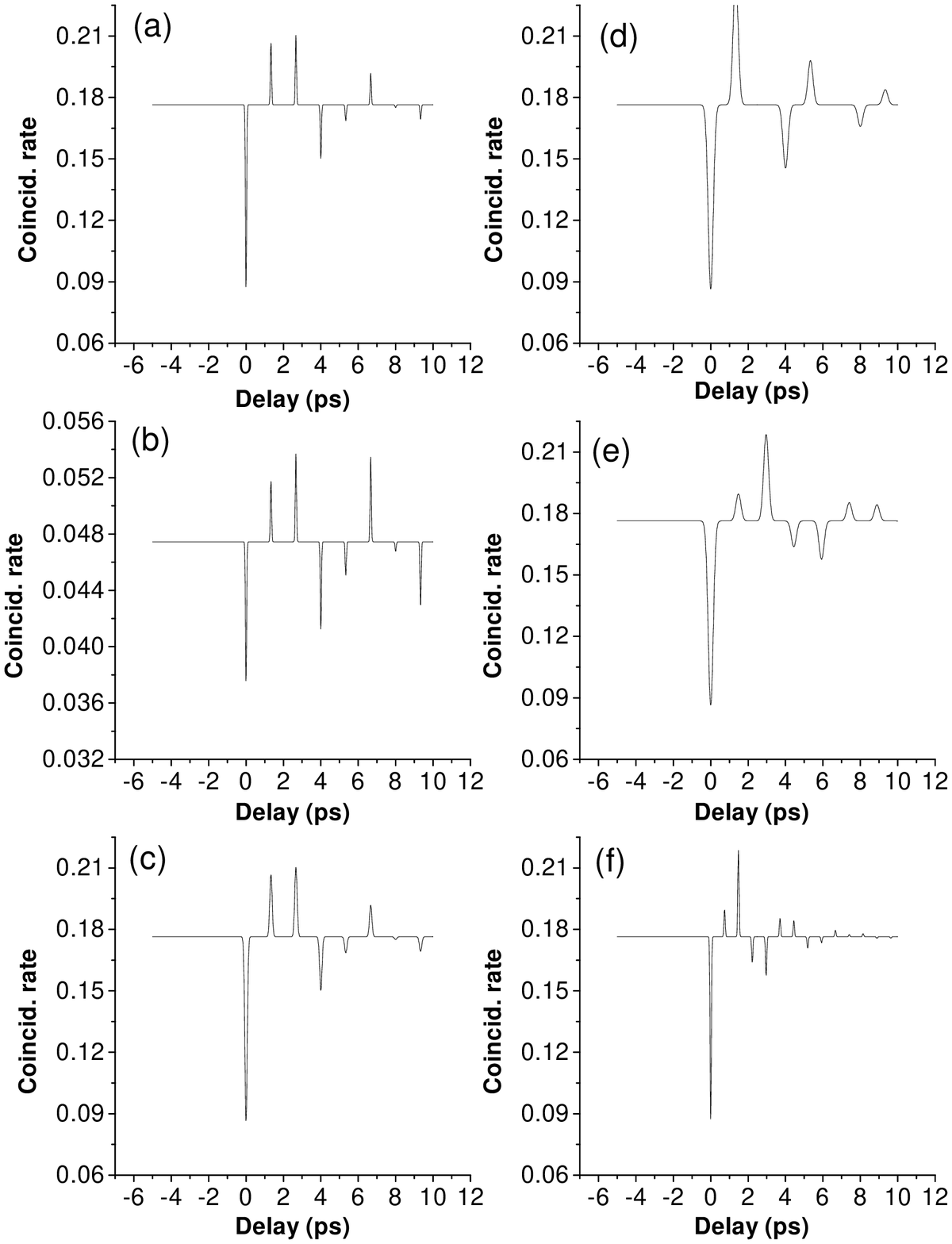}}
\caption{We begin in (a) with $\lambda=826.2$ nm, $\Delta\lambda=8$ nm, $L=0.4$ mm, $R=0.7$.
In (b) we change only $R$ to 0.9. In (c) we come back $R$ to 0.7, maintaining all the other
parameters the same, and change $\Delta \lambda$ to $4$ nm. In (d), in relation to (c),
 we only change $\lambda$ to $1200$ nm. In (e), we made $L$ equal to $0.44444$ mm.
 Finally in (f), we change $\lambda$ to $600$ nm and $L$ to $0.22222$ mm.} \label{ucvp}
\end{figure}

Figures (\ref{ucvr}), (\ref{ucar}) and (\ref{ucvp}) show the
behavior of the number of coincidences, according to Eq.
(\ref{equ1}), as a function of the delay $\delta$. The values for
the different parameters entering in this equation were taken from
the experimental data reported by Sagioro et al \cite{Sagioro}. In
Fig. \thinspace(\ref{ucvr}) the length $L$ of the cavity was
chosen as an integer multiple of half of the central wavelength of
the photon wavepacket. In this case, the curve is composed of a
mesa function intersected by equally separated valleys. In
Fig.\thinspace(\ref{ucar}) the length of the cavity was chosen as
half integer multiple of half of the central wavelength of the
photon wavepacket. In this case the curve is composed of a mesa
function intersected by alternated valleys and peaks. We refer to
these cases as resonant and anti-resonant respectively. A
different situation is depicted in Fig.\thinspace(\ref{ucvp}a)
which shows the number of coincidences when the conditions for
resonance or anti-resonance are not fulfilled. In this particular
case peaks and valleys do not alternate following a regular
pattern. Fig.\thinspace(\ref{ucvp}) also shows the behavior of the
number of coincidences when we change the value of the other
parameters entering in Eq.\thinspace(\ref{equ1}). In
Fig.\thinspace (\ref{ucvp}b) the value of $R$ has been increased
from $0.7$ to $0.9$. The only noticeable effect is a change in the
value of the mesa function and on the amplitudes of valleys and
peaks. The positions of peaks and valleys remains unchanged. In
Fig.\thinspace(\ref{ucvp}c) we change $\bigtriangleup\lambda$ from
8 nm to 4 nm leaving the other parameters as in
Fig.\thinspace(\ref{ucvp}b). In this case increases the width of
valleys and peaks leaving the other features of the number of
coincidences unchanged. Figure (\ref{ucvp}d) shows the
modification of the number of coincidences when the wavelength is
changed from 826.2 nm to 1200 nm. Notice that some of the valleys
and peaks appears to be absent. In Fig.\thinspace(\ref{ucvp}e) the
parameters are as in Fig.\thinspace(\ref{ucvp}c) with $L=0.44444$.
Finally, in Fig.\thinspace(\ref{ucvp}f) $L$ and $\lambda$ were
changed while keeping the ratio $L/\lambda$ fix. In this case the
number of coincidences behaves like in
Fig.\thinspace(\ref{ucvp}e).

\begin{figure}[h]
\centerline{\includegraphics[angle=0,width=8cm]{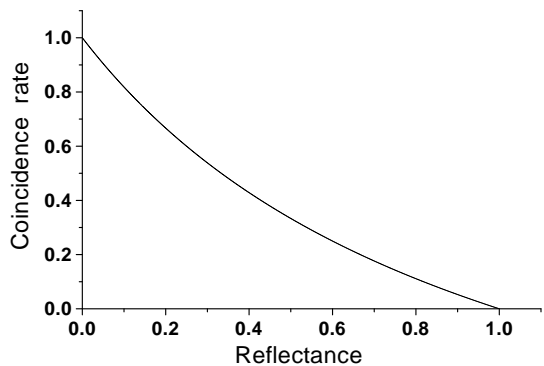}}
\caption{Graphic showing the coincidence rate versus reflectance $(R)$. We put $L=0.4$ mm,
 the delay in $0.66733$ ps, which is approximately in the middle between the first and the
 second interferences (platform); $\lambda= 826.2$ nm and $\Delta\lambda=8$ nm.} \label{ucvR}
\end{figure}

The value of the mesa function can be obtained easily. We consider $N_c$ as a function of
the delay $\delta$. Thus, the first term entering in Eq.\thinspace(\ref{equ1}) turns out to
 be a constant. Since, for the values of the parameters that we consider, the ratio
  $2\pi\bigtriangleup\lambda L/\lambda^2$ is larger than $30$ the exponential function
  in the first term of Eq.\thinspace(\ref{equ1}) approximately vanishes for values of
  $n-q\neq 0$. Thereby, Eq.\thinspace(\ref{equ1}) becomes
\begin{equation}
N_c= \frac{T^2}{1-R^2} - T^2\sum_{n,q} R^{n+q} \cos\big(\frac{2\pi
L}{\lambda_p}(n-q)\big) e^{-(\frac{2\pi c \bigtriangleup\lambda
}{\lambda^2})^2 (\frac{L}{c}(n+q)-\delta)^2}. \label{equ2}
\end{equation}
Figure (\ref{ucvR}) shows the number of coincidences as a function of the reflectance $R$.
 The curve coincides exactly with the function $T^2/(1-R^2)$. Let us see where this function
 comes from. The probability of one photon to be transmitted through a cavity is $T$ and to
 be reflected is $R$. Summing over all the possibilities (a photon crossing the cavity
 directly, a second reflecting inside twice before crossing it, a third reflecting four
 times and so on), and obtain the total probability for crossing the photons $(P)$:
$P=T^2+T^2R^2+T^2R^4+T^2R^6+...=\frac{T^2}{1-R^2}$. This result shows that, when the
longitudinal coherence length $\lambda^2/\bigtriangleup\lambda$ of the single photons
is much smaller than the cavity length $L$, the photon acts like particles with probability
 $T$ for crossing an obstacle and probability $R$ to be reflected, i.e. there is no
 interference between a photon with itself, as there was in a traditional Fabry-Perot cavity.

\begin{figure}[h]
\centerline{\includegraphics[angle=0,width=7cm]{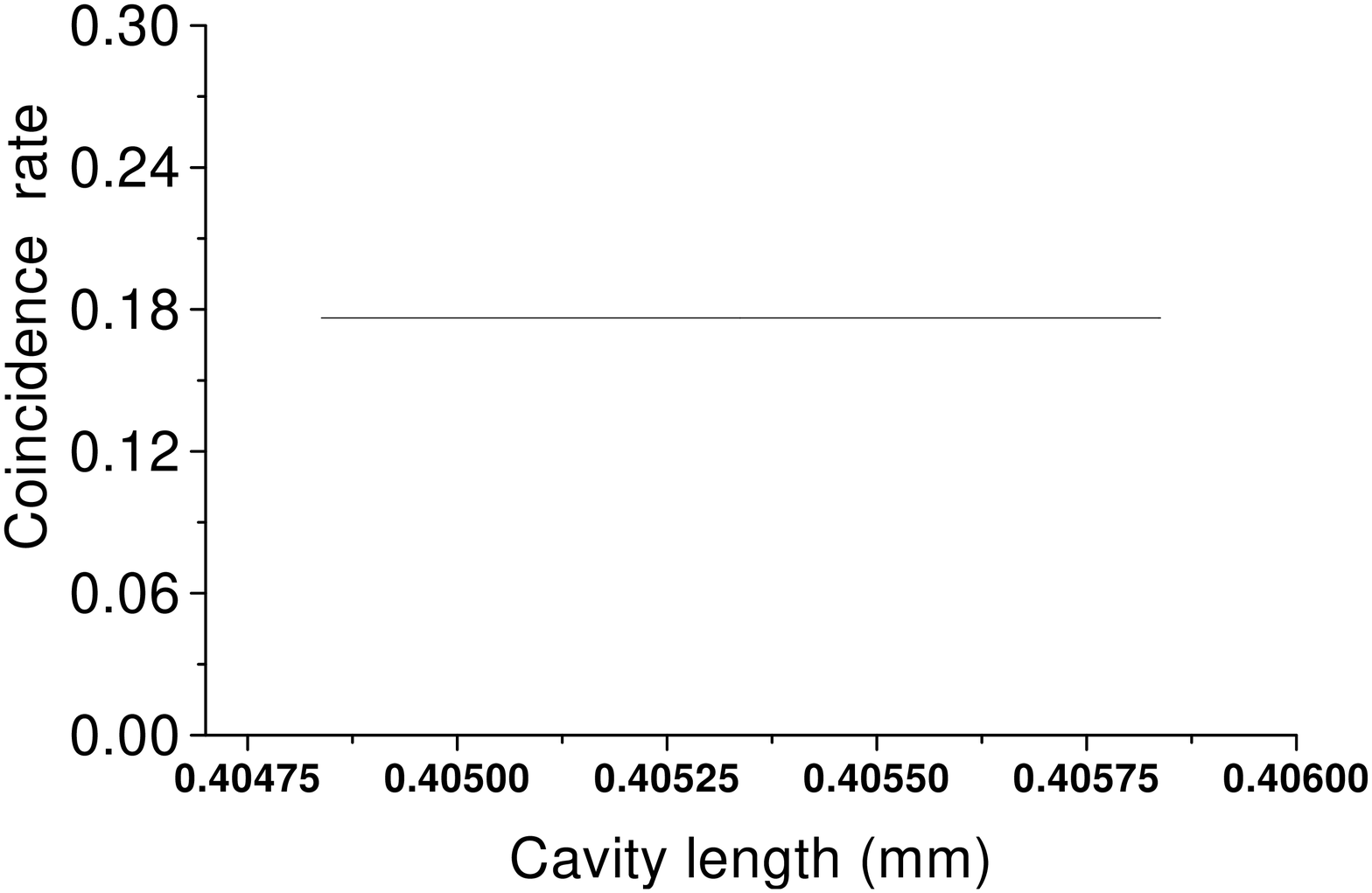}}
\caption{Graphic showing coincidence rate versus cavity length. We begin with $L=0.404838$
 mm and fix the delay in $0.66733$ ps, which is approximately in the middle between the
  first and the second interferences (platform), and vary $L$ just a few lambdas.
  The parameters are $R=0.7$, $\lambda=826.2$ nm and $\Delta\lambda=8$ nm.} \label{ucvl}
\end{figure}

\begin{figure}[h]
\centerline{\includegraphics[angle=-90,width=14cm]{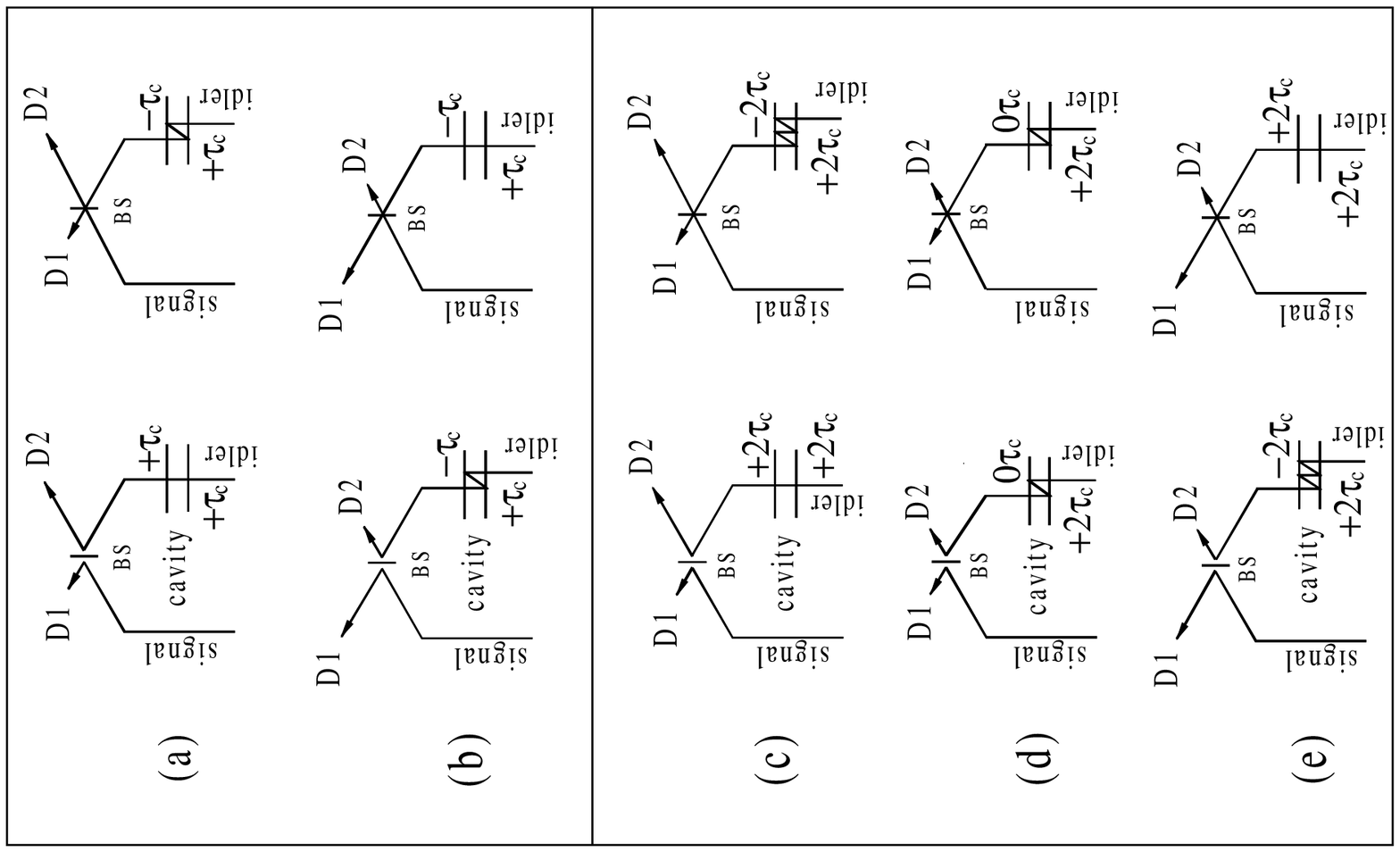}}
\caption{(a) and (b) are Feynman's path diagrams associated with the second interference
region ($\delta=\tau_{c}$). (c), (d) and (e) are  Feynman's path diagrams associated with
the third interference region ($\delta=2\tau_{c}$).} \label{ucd}
\end{figure}

The second term in Eq. (\ref{equ1}) composes of a sum over
unnormalized gaussian functions
$e^{-(\delta-\tilde\delta)^2/2\sigma^2}$ centered at
$\tilde\delta_{n+q}=(L/c)(n+q)$ with standard deviation
$\sigma=\lambda^2/(2\sqrt{2}\pi c \bigtriangleup\lambda)$, which
is independent of $n$ and $q$. Let us note that the distance
between the centers of two consecutive gaussian functions is given
by $L/c$. The overlap between two consecutive gaussian functions
is given by
\begin{equation}
\int_{-\infty}^{+\infty}d\delta
e^{-(\frac{\delta-\tilde\delta_{n+q}}{\sqrt{2}\sigma})^2}
e^{-(\frac{\delta-\tilde\delta_{n+q+1}}{\sqrt{2}\sigma})^2}
=
\sqrt{\pi}\sigma
e^{-(\frac{\tilde\delta_{n+q}-\tilde\delta_{n+q+1}}{2\sigma})^{2}}
\end{equation}
being its value tenth of picoseconds for the value of the parameters here considered.
Thus, it is possible to distinguish consecutive gaussian functions. These gaussian functions
lead to the peaks and valleys in figures (\ref{ucvr}), (\ref{ucar}) and (\ref{ucvp}).
Peaks or valleys appear for values of $\delta$ fulfilling the condition
\begin{equation}
\delta=\frac{L}{c}(n+q)
\end{equation}
with a maximum amplitude given by
\begin{equation}
N_c(j)=\frac{T^2}{1-R^2} - T^2\sum_{n,q}R^{n+q}\cos\big(\frac{2\pi
L}{\lambda_p}(n-q)\big). \label{amplitud}
\end{equation}
where $j$ is the order of the peak or valley ($j=1$ for the first, in $\delta=0$; $j=2$
for the second, in $\delta=\tau_c$; $j=3$ for the third, in $\delta=2\tau_c$ and so on)
 and the summation is for all $n+q=j-1$.

Let us now consider for instance the condition for the resonant
case. In this case, the cosine function entering in Eq.
(\ref{amplitud}) is always one. Thereby, the amplitude of the
valleys becomes
\begin{equation}
N_c(j)=\frac{T^2}{1-R^2} - T^2R^{j-1}j.
\end{equation}
In the anti-resonant case the amplitude of the peaks and valleys
is given by
\begin{equation}
N_c(j)=\frac{T^2}{1-R^2} - T^2(-R)^{j-1}j.
\end{equation}

The origin of peaks and valleys can be illustrated by considering the diagrams depicted in
Fig. (\ref{ucd}). Fig. (\ref{ucd}a) shows the signal beam with a delay $\delta=\tau_{c}$
which is produced by changing the length of the optical path (see Fig. (\ref{hom2})).
The photons of the idler beam are transmitted through the cavity without being reflected by
the mirrors of the cavity. Both beams are reflected at the beam splitter. The overall effect
 is that detector D2 detects a photon a time $\tau_{c}$ before detector D1. The situation
 depicted in Fig. (\ref{ucd}a') is slightly different but the overall effect remains
 unchanged. In this case photons of the idler beam reflect twice inside the cavity but
 photons of both beams are transmitted at the beam splitter, so that detector D2 detects
 the photons a time $\tau_{c}$ before detector D1. The situation is similar in the cases
  of Fig. (\ref{ucd}b) and (\ref{ucd}b'), the only difference is that detector D1 detects
  photons a time $\tau_{c}$ before detector D2. These four cases, which are partially
  distinguishable due to their different probabilities, contribute to the same peak or
  valley at $\delta=\tau_{c}$ in the coincidence detection. The interference at
  $\delta=2\tau_{c}$ can also be explained in these terms. Figures (\ref{ucd}c),
  (\ref{ucd}c'), (\ref{ucd}d), (\ref{ucd}d'), (\ref{ucd}e) and (\ref{ucd}e') show all
  the possible processes contributing in this case. Notice that in the process depicted
  in Fig. (\ref{ucd}d) the photons overlap at the beam splitter, but in the other processes
  there is no such an overlap between the photons. When $\delta=0$, the only possible
  process is the overlap between the photons at the beam splitter caused by the direct
  crossing of photons of the idler beam through the cavity.

Figure \ref{ucd} shows why the interferences always happen when $\delta$ is a multiple
of $\tau_{c}$. By analyzing  Eq. (\ref{equ}) we deduce the coincidence rate value for each
interference region:
\begin{equation}
N_c(j)=\frac{T^2}{1-R^2} - T^2\sum_{n,q}R^{n+q}\cos{[\omega_p\tau_c(n-q)]}.
\end{equation}
Where $j$ is the order of the interference region ($j=1$ for the first, in $\delta=0$;
 $j=2$ for the second, in $\delta=\tau_c$; $j=3$ for the third, in $\delta=2\tau_c$ and so on)
  and the summation is for all $n+q=j-1$.

Now we would like to come back to Fig. \ref{ucvr}, remember that increasing $R$ the deepest
valley is each time more distant from the origin, we will give a brief explication for
such behavior. Look at Fig. \ref{ucd}, parts (a) and (b) are associated with the second
interference region ($\delta=\tau_{c}$), parts (c), (d) and (e) are associated with the
third one ($\delta=2\tau_{c}$). We could write out parts (f), (g), (h) and (i) associated
with the fourth interference region, if we want to, and so on. Notice that increasing the
 order of the interference region, the number of pairs of interfering processes increases
 the same quantity. On the other hand, the probability of occurrence of these processes
  decreases. The case shown in Fig. \ref{ucvr} is the resonant one, when all the interference
   contributions are destructive. To see this, take Eq. ( \ref{equ}) and make
   $\omega_{p}\tau_{c}=E\pi$, where E is an even number (resonance condition).
   By analyzing the same Eq. (\ref{equ}), we conclude that the probability of occurring
   determinate pair of interference process is proportional to $R^{(j-1)}$,
   where $R$ is the cavity mirror reflectance and $j$ is the order of the interference region.
    The number of pairs of interference processes in each interference region is $j$,
     so the total probability of occurring determinate interference region is proportional
     to $jR^{(j-1)}$, $P(j,R) \propto jR^{(j-1)}$. The highest $P(j,R)$ corresponds to the
     deepest valley (in the resonant case). Once we have $P(j,R)$ it is easy to show that
     its highest value, when you increase $R$ is for a higher $j$. This highest value for
     $P(j,R)$ when $R\rightarrow1$ is when $j\rightarrow\infty$.

%Simulaciones con cavidades simitricas cambiando reflectividad del %espejo, filtro de interferencia, resonancia y antiresonancia de la %cavidad (bunching y anti-bunching). Discusisn sobre las %visibilidades. Explicacisn de los diagramas de Feymann.

\section{HOM with two cavities}

Now we will show some simulations using the complete Eq.
(\ref{eqg}) with the constant $\pi K$ removed. Figure
\ref{dc,rr,rar,arar} shows the interference patterns when two
cavities are present in the interferometer.

\begin{figure}[h]
\centerline{\includegraphics[angle=0,width=8cm]{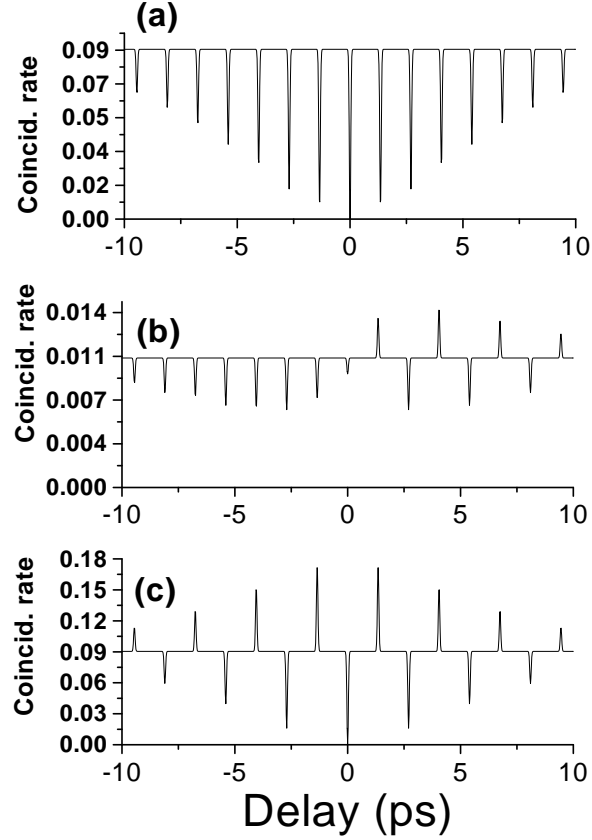}}
\caption{In this graphic both cavities have the parameters: $R_{i}=R_{s}=0.7,
 \lambda_{i}=\lambda_{s}=826.2$ nm and $\Delta\lambda=8$ nm. In (a)
  $L_{i}=L_{s}=0.404838$ mm, both cavities are resonant with the twin photons.
  In (b) $L_{s}=0.404838$ mm (resonant), $L_{i}=0.4050447$ mm (anti-resonant).
  In (c) $L_{s}=L_{i}=0.4050447$ mm, both cavities are anti-resonant.}
   \label{dc,rr,rar,arar}
\end{figure}

Notice that in Fig. \ref{dc,rr,rar,arar}.a and in Fig. \ref{dc,rr,rar,arar}.c we have
symmetric patterns with respect to the origin, but we do not have this symmetry in
Fig. \ref{dc,rr,rar,arar}.b. Notice also that there are interferences for negative delays,
 not only a platform as in the one cavity problem (Fig. \ref{ucvr}-\ref{ucvp}).
  Finally notice that in Fig. \ref{dc,rr,rar,arar}.a and Fig. \ref{dc,rr,rar,arar}.c,
  when $\delta=0$ we get 100\% of coalescence. This totally destructive interference is
  not possible, when the longitudinal coherence of the photons is much smaller than the
  cavity length, using only one cavity.

\begin{figure}[h]
\centerline{\includegraphics[angle=0,width=8cm]{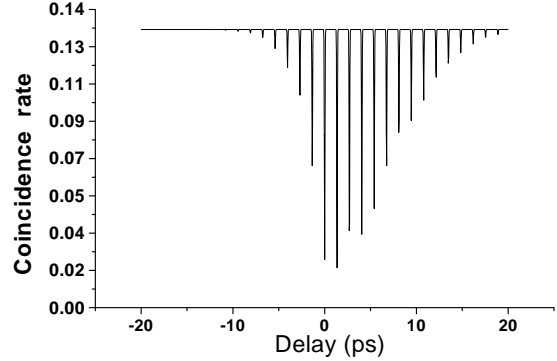}}
\caption{The parameters are $\lambda=826.2$ nm, $\Delta\lambda=8$ nm,
$L_{i}=L_{s}=0.404838$ mm (both resonant), $R_{i}=0.7$ and $R_{s}=0.4$. } \label{dcd}
\end{figure}

Let us see what happens when the cavities are unbalanced (Fig. \ref{dcd}).

In Fig. \ref{dcd} the coincidence pattern is composed only of valleys (both cavities are
 resonant), but the platform changes its value from what it was in
 Fig. \ref{dc,rr,rar,arar}.a because the percentages of crossing photons are different
 ($R_{i}=0.7$ and $R_{s}=0.4$).

\begin{figure}[h]
\centerline{\includegraphics[angle=0,width=8cm]{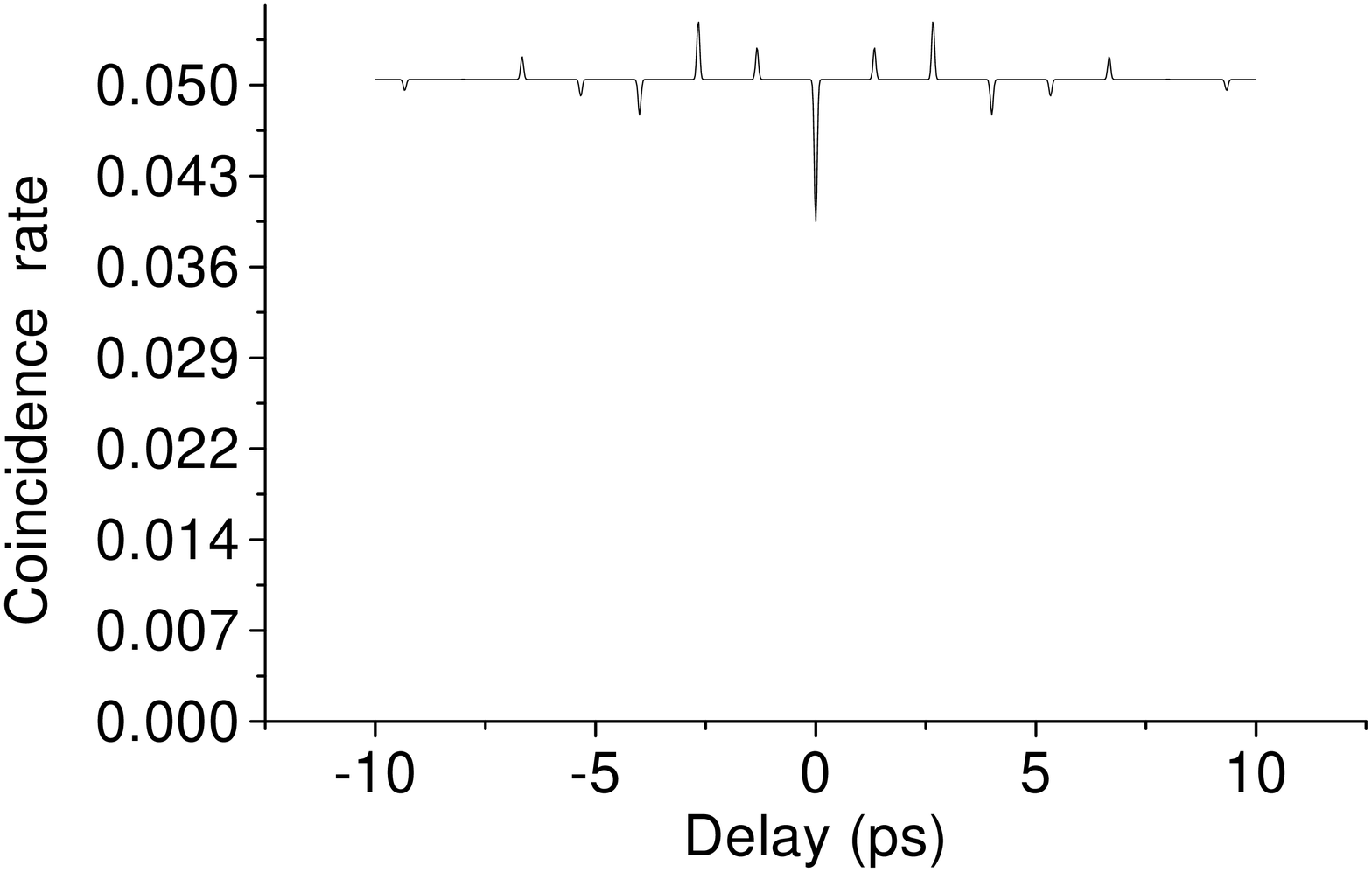}}
\caption{The parameters are $\lambda=826.2$ nm, $\Delta\lambda=8$ nm, $R_{i}=R_{s}=0.7$
 and $L_{i}=L_{s}=0.4$ mm (neither resonant nor anti-resonant).} \label{dcl04}
\end{figure}

Now we take the case of identical cavities without resonances and anti-resonances
(Fig. \ref{dcl04}).

See in Fig. \ref{dcl04} that the coincidence pattern is symmetric with respect to the
 origin, because $L_{i}=L_{s}$, but at this time the coincidence rate platform has a
 lower value than in Fig. \ref{dc,rr,rar,arar}.a and Fig. \ref{dc,rr,rar,arar}.c,
  because the cavities are neither resonant nor anti-resonant.

\begin{figure}[h]
\centerline{\includegraphics[angle=0,width=8cm]{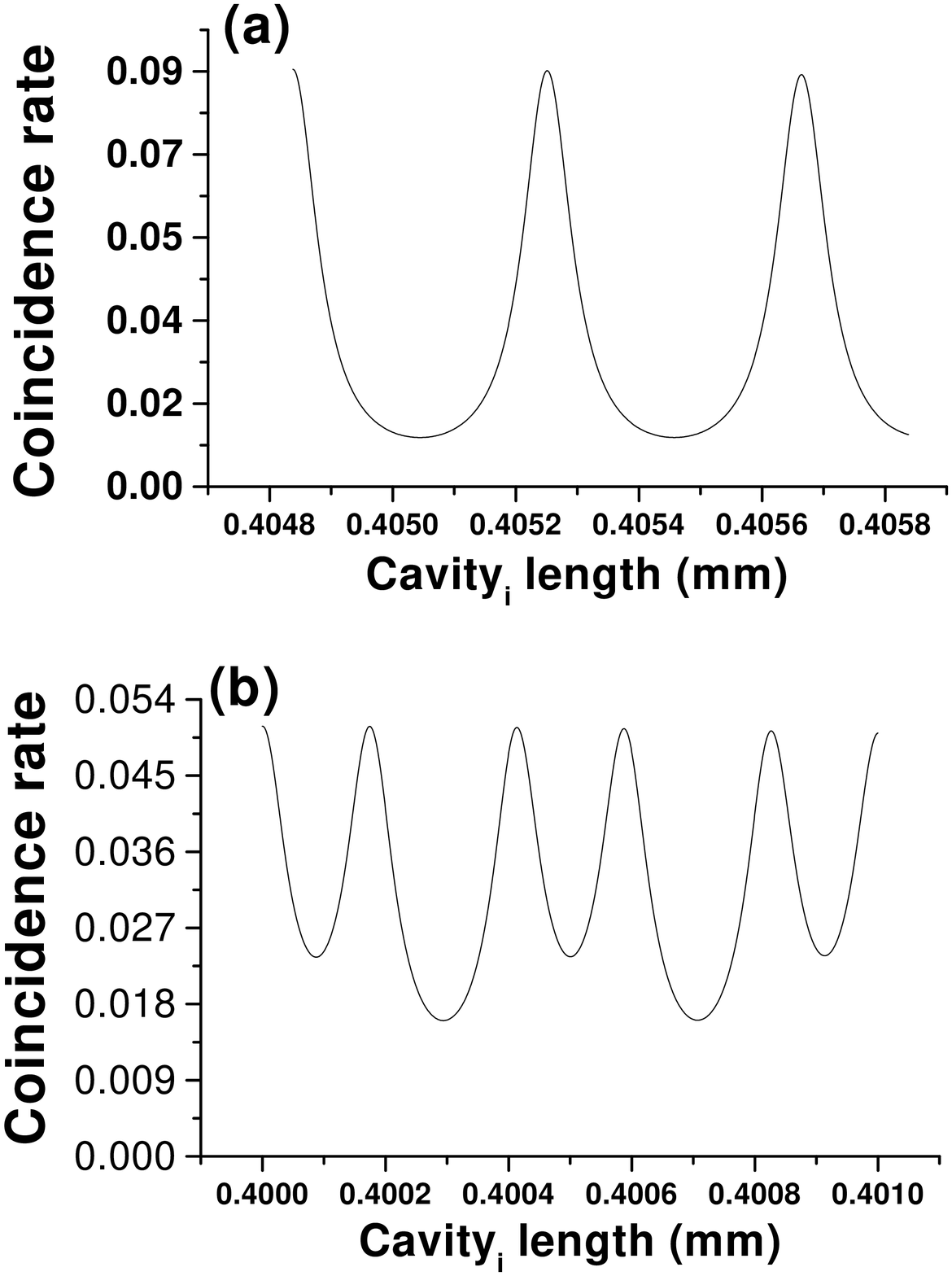}}
\caption{Graphics showing coincidence rate versus $L_{i}$. All the other parameters are
kept fix: $R_{i}=R_{s}=0.7$, $\lambda=826.2$ nm, $\Delta\lambda=8$ nm, $delay=0.66733$ ps
(platform), in (a) $L_{s}=0.404838$ mm (resonant), in (b) $L_{s}=0.4$ mm
(neither resonant nor anti-resonant).} \label{dcvl}
\end{figure}

In the one cavity case, the platform does not depend on $L$ and has an special meaning
(Fig. \ref{ucvl}). How would this be with two cavities? See Fig. \ref{dcvl}.

In Fig. \ref{dcvl}.a we begin with $L_{i}=0.404838$ mm (resonant) and vary it just a few
lambdas, passing by anti-resonances, intermediate cases, coming back to resonances and so on.
Notice now that, unlike Fig. \ref{ucvl}, the platform depends on $L$. Its value is maximum
exactly when $L_{i}$ is resonant (because $L_{s}$ is resonant too) and is minimum when
$L_{i}$ is anti-resonant. With two cavities the platform does not mean the percentage of
 crossing photons anymore, i.e. you never are free of interferences, any place you put
  the delay, including the platform, you have some kind of interference (this will be better
   understood when we study the paths interference Feynman's diagrams for the two cavities).
    The distance between consecutive maximums is $\lambda/2$. In Fig. \ref{dcvl}.b, the
    only difference from Fig. \ref{dcvl}.a is that we fixed $L_{s}$ in $L_{s}=0.4$ mm
    (neither resonant nor anti-resonant) and began with $L_{i}=0.4$ mm. The procedure was
    the same (to vary $L_{i}$ just a few lambdas). Now the distance between consecutive
     minimums is $\lambda/4$.

\begin{figure}[h]
\centerline{\includegraphics[angle=-90,width=8cm]{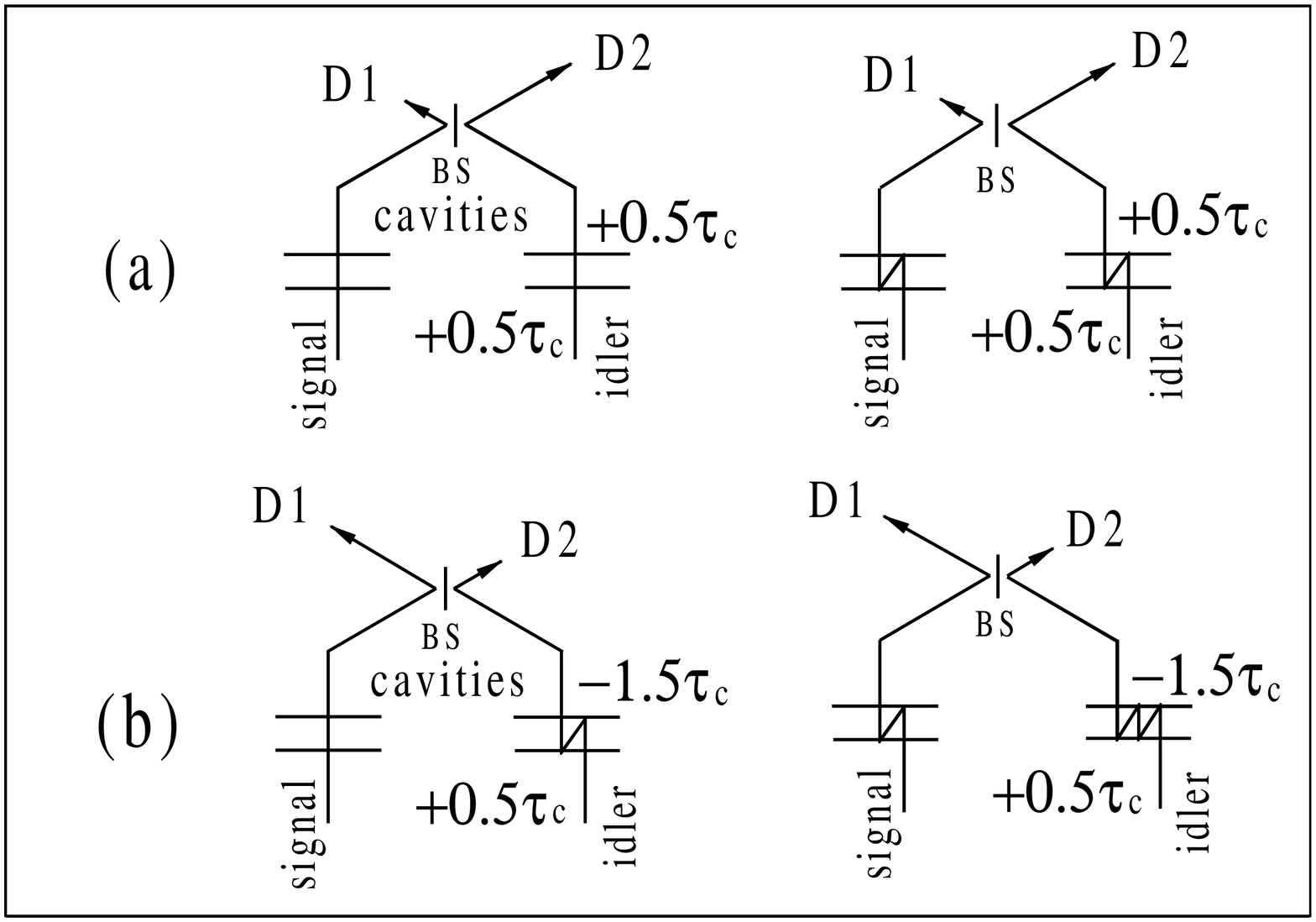}}
\caption{Paths interference Feynman's diagrams for two cavities, explaining the platform
variation with $L$. Photon idler begin $0.5\tau_{c}$ advanced in relation to photon signal.
In (a) it remains $0.5\tau_{c}$ advanced, after both twin photons cross their cavities.
In (b), photon signal, after both photons cross their cavities, is advanced $1.5\tau_{c}$
in relation to photon idler.} \label{dcdiag}
\end{figure}

We can explain the platform variation with $L$ by analyzing Fig. \ref{dcdiag}.

The reasons for the presence of peaks and dips in the interference pattern of the HOM
interferometer with two cavities are the same presented in Fig. \ref{ucd}: you do not know
what photon arrives in a detector first. But the coincidence rate platform with just one
cavity does not depend on $L$ (Fig. \ref{ucvl}). When you have two cavities, the platform
itself depends on $L$ (Fig. \ref{dcvl}). We could explain this dependence in a general
 fashion, but, to make the explication easier to understand, we took a particular case.
 In Fig. \ref{dcdiag}, $\tau_{cs}\simeq\tau_{ci}$ or $\tau_{cs}=\tau_{ci}$, so we call
 them $\tau_{c}$. We began with photon idler advanced by $0.5\tau_{c}$ with respect to
 photon signal (we made use of prisms P1 and P2 to do so, see Fig. \ref{hom2}).
 In Fig. \ref{dcdiag}.a, both photons can cross its cavities directly or both can reflect
  inside the cavities twice before crossing them or (this is not shown here) they can
   reflect inside their cavities four times before crossing the cavities and so on;
   in all these cases photon idler still would be $0.5\tau_{c}$ advanced. In
   Fig. \ref{dcdiag}.b, because of the different amounts of reflections between the
   photons inside their cavities, photon signal arrives $1.5\tau_{c}$ before photon idler
    in BS, and this can happen by various different ways. If the difference between the
     clicks at the detectors is $0.5\tau_{c}, 2.5\tau_{c}, 4.5\tau_{c}$ and so on, you
      know that was photon idler that arrived first, if the difference is $1.5\tau_{c},
      3.5\tau_{c}, 5.5\tau_{c}$ and so on, you know that was photon signal that arrived first.
       Unlike the interferences that causes the peaks and dips, when you do not know with
       certainty what photon arrives first at a detector, in this kind of interference
       (that makes the platform vary) you know what photon arrives first, but you do not
        know which process has happened before the arriving. Resuming: in Fig. \ref{dcdiag},
         all process in line (a) interfere, all process in line (b) interfere and so on,
          this is the very reason for the fluctuation of the platform.

 \section{Applications: control not gate}

%Estudiar csmo se modifican el nzmero de coincidencias con respecto
%al caso de una cavidad. XOR-Clasica.
This setup (Fig. \ref{hom2}) is suitable for applications in information
problems. For example: we can construct an optical $XOR$
gate \cite{chuang}, in a similar way as shown in ref.
\cite{becker}. In their case the control was made by changing
electro-optically the birefringence of two quartz plates with
different lengths inserted at each arm of the HOM. In our case the
control is made by changing the length of the cavities. The HOM
interference pattern changes if the length of each cavity is an
integer multiple of half wavelength (``resonance'') or a
semi-integer multiple of half wavelength (``anti-resonance''). The
logical bits are encoded in the resonances/antiresonances of the
cavities with the photon central wavelength similarly to the
phases of photon 1 and 2 in the setup of Becker et al.
\cite{becker}. Calculations concerning two cavities, each one in
one of the arms of the interferometer, showed above the behavior
of the coincidence counts. Consider that the cavities are nearly
equal in length. If both central wavelength of idler and signal
photons are resonant or anti-resonant with their cavities, you
have a symmetric interference pattern, in relation to $\delta=0$, in the coincidence counts.
 But if one of the photons is
resonant with its cavity and the other is anti-resonant with its
one, it is obtained a not symmetric coincidence pattern (in
relation to $\delta=0$), see Fig. \ref{dc,rr,rar,arar}. Now let us identify input bit 0
with the resonance and input bit 1 with the anti-resonance. We
also identify output bit 0 with the symmetric pattern in the
coincidences and output bit 1 with the not symmetric pattern. With
these identifications, it is easy to simulate a logical XOR
operation, see table \ref{XOR}.
\begin{table}[h]
\begin{center}
\begin{tabular}{|c|c|c|}
\hline
\multicolumn{3}{c}{\textbf{Table for XOR gate}} \\
  \hline
  Idler & Signal & Results\\
  bit cavity& bit cavity& pattern  bit\\
  \hline
  0 \,\,\,\,\,\,\,\, res& 0 \,\,\,\,\,\,\,\, res & SY \,\,\,\,\,\,\,\, 0\\
  0 \,\,\,\,\,\,\,\,res& 1 \,\,\,\,\,\,\,\,ant & NS \,\,\,\,\,\,\,\, 1\\
  1 \,\,\,\,\,\,\,\,ant& 0 \,\,\,\,\,\,\,\,res& NS \,\,\,\,\,\,\,\, 1\\
  1 \,\,\,\,\,\,\,\,ant& 1 \,\,\,\,\,\,\,\,ant & SY \,\,\,\,\,\,\,\, 0\\
  \hline
\end{tabular}
\caption{Table showing the behavior of a logical XOR operation.
``res'' stands for resonant, ``ant'' for anti-resonant, ``SY''
stands for the symmetric coincidence pattern and ``NS'' for the
not symmetric one.} \label{XOR}
\end{center}
\end{table}

Table \ref{XOR} shows that we can choose photon idler as the
control and photon signal as the target, for simulating the XOR
gate, or vice-versa. This interchanging between control and target
is very useful in constructing logical circuits.

\section{Conclusion}

We started making the calculations for the coincidence measures in the Hong-Ou-Mandel
interferometer with two symmetric cavities, each one put in one arm of the interferometer.
Using the resulting equation we made some simulations. It was easy to reduce the general
 equation for two cavities to the more specific one for just one cavity. One interesting
 result in this case (one cavity) is the observation of coalescence and anti-coalescence,
  both comportment are determinate by the rate between the cavity length and the wavelength
   of the photon that cross it, although the photon coherence length is smaller than the
    cavity length. But, the main conclusion concerning anti-coalescence is that this is
     only possible due to the non-overlapping processes happening from the second
     interference region (see Fig. \ref{ucd}.a,b) in ahead. If the photons overlap in the
     beam splitter (having the same polarization, frequency and transverse mode) they
     necessarily leave the beam splitter by the same output, giving origin to no coincidence.
      In the two cavities case, there are two results that deserve to be pointing.
       The first is the null result for the coincidences when
       $\delta=0$ (Fig. \ref{dc,rr,rar,arar}.a,c), this was not possible in the one cavity
       case. The second is the variation of the platform with the changing in the cavities
        lengths, this is due a new kind of interference that either did not happen in the
        one cavity case. Finally we show that is possible to construct a XOR gate using our
         apparatus. In fact, if you prefer, instead of using as outgoing bits, SY standing
         for bit 0 and NS for bit 1 (see table \ref{XOR} last collum) you can use the value
         of the platform (see Fig. \ref{dc,rr,rar,arar}), all you have to do is to stop the
         delay in the platform, e.g. $\delta=0.66733$ ps in that case, calibrate your system,
          and you are ready to use the platform value as outgoing bits, for example:
          the highest value can stand for bit 0 and the lowest one for bit 1.

\section{Acknowledgements}

This work was supported by CNPq, Pronex-Semicondutores and
Mil\^{e}nio-Informa\c{c}\~{a}o Qu\^{a}ntica. We would like to
thank S. P. Walborn for pointing us ref. \cite{becker}. A.D. was
supported by Fundaci\'on Andes.

\end{document}